\documentclass[]{spie}  

\usepackage{amsmath,amsfonts,amssymb, mathrsfs}
\usepackage{bigints}
\usepackage{subfig}
\usepackage{graphicx}
\usepackage[colorlinks=true, allcolors=blue]{hyperref}
\usepackage{stmaryrd}

\usepackage[braket,qm]{qcircuit}
\pdfmapfile{+sansmathaccent.map}

\usepackage{braket}
\usepackage{physics}

\usepackage{xcolor}
\usepackage{listings}
\usepackage{color}

\definecolor{codegreen}{rgb}{0,0.6,0}
\definecolor{codegray}{rgb}{0.5,0.5,0.5}
\definecolor{codepurple}{rgb}{0.58,0,0.82}
\definecolor{backcolour}{rgb}{0.95,0.95,0.92}
 
\lstdefinestyle{mystyle}{
    backgroundcolor=\color{backcolour},   
    commentstyle=\color{codegreen},
    keywordstyle=\color{magenta},
    numberstyle=\tiny\color{codegray},
    stringstyle=\color{codepurple},
    basicstyle=\footnotesize,
    breakatwhitespace=false,         
    breaklines=true,                 
    captionpos=b,                    
    keepspaces=true,                 
    numbers=left,                    
    numbersep=5pt,                  
    showspaces=false,                
    showstringspaces=false,
    showtabs=false,                  
    tabsize=2
}
\usepackage{todonotes}

\DeclareSymbolFont{largesymbolsA}{U}{txexa}{m}{n}
\DeclareMathSymbol{\varprod}{\mathop}{largesymbolsA}{16}

\title{Practical numerical integration on NISQ devices}

\author[a,*]{Kwangmin Yu}
\author[b]{Hyunkyung Lim}
\author[c]{Pooja Rao}
\affil[a]{Computational Science Initiative, Brookhaven National Laboratory, Upton, New York 11973, USA}
\affil[b]{Department of Applied Mathematics and Statistics, Stony Brook University, Stony Brook, New York 11794, USA}
\affil[c]{Department of Mathematics, Stony Brook University, Stony Brook, New York 11794, USA}

\authorinfo{* : Send correspondence to Kwangmin Yu,  E-mail : kyu@bnl.gov\\
}

\begin{document} 
\maketitle

\begin{abstract}
This paper addresses the practical aspects of quantum algorithms used in numerical integration, specifically their implementation on Noisy Intermediate-Scale Quantum (NISQ) devices. Quantum algorithms for numerical integration utilize Quantum Amplitude Estimation (QAE) (Brassard et al., 2002) in conjunction with Grover’s algorithm. However, QAE is daunting to implement on NISQ devices since it typically relies on Quantum Phase Estimation (QPE), which requires many ancilla qubits and controlled operations. To mitigate these challenges, a recently published QAE algorithm (Suzuki et al., 2020), which does not rely on QPE, requires a much smaller number of controlled operations and does not require ancilla qubits. We implement this new algorithm for numerical integration on IBM quantum devices using Qiskit and optimize the circuit on each target device. We discuss the application of this algorithm on two qubits and its scalability to more than two qubits on NISQ devices.   
\end{abstract}


\section{Introduction}

Since quantum computing (QC) was first introduced in the 1980s \cite{feynman1982simulation},
it has grown into an active and diverse field of research.
In recent years, significant progress has been made in building quantum computers by companies, such as IBM, Google, and Rigetti. 
Recently, Google, IBM, and Intel have announced 72 qubits \cite{Kelly2018Google}, 50 qubits \cite{Knight2017IBM} (53 qubits \cite{SHANKLAND2019IBM}), and 49 qubits \cite{Hsu2019Intel} quantum devices, respectively. 
More notably, IBM has made available its cloud enabled quantum computing platform to the public.  
These currently available NISQ devices provide a tangible quantum programming environment and are serving as a stepping stone for the large-scale universal quantum computers of the future \cite{Preskill2018quantumcomputingin}. 
Despite the significant breakthrough achieved in qubit numbers in the NISQ era, the quantum systems are limited by (a) the short coherence times, the amount of useful operational time in a calculation before the information loss, (b) the circuit depth, the number of sequential quantum operations that can be performed on a quantum device, and (c) the lack of error correction \cite{Preskill2018quantumcomputingin}. They are so noisy that the number of quantum gates that can be implemented is significantly impacted \cite{tannu2019not, corcoles2019challenges}. Furthermore, due to limited physical inter-connectivity between the qubits, the multi-qubit gates (operations), such as controlled-NOT and controlled-controlled-NOT (Toffoli) are hard to implement efficiently~\cite{tannu2019not, Adbo2019ibmq}.

The aforementioned device limitations restrict the implementation of the quantum algorithms developed in the 1990s such as Quantum Fourier Transform (QFT) \cite{Shor:1997:PAP:264393.264406} and Quantum Phase Estimation (QPE)\cite{kitaev1995quantum}. In particular, QPE, which requires many controlled operations and ancilla qubits, plays a critical role in various quantum algorithms, including Shor's factoring algorithm\cite{Shor:1997:PAP:264393.264406}. This has motivated several efforts to design a more efficient version of QPE \cite{Svore:2014:FPE:2600508.2600515, berg2019practical}, but the number of controlled operations is still restrictive. Therefore, the Quantum Amplitude Estimation (QAE) of Brassard et al. (canonical QAE) \cite{brassard2002quantum}, which relies heavily on QPE, is daunting to implement on NISQ devices.

Several variants, such as maximum likelihood quantum amplitude estimation (MLQAE), iterative quantum amplitude estimation (IQAE), etc., have recently been published \cite{aaronson2020quantum, suzuki2020amplitude, grinko2019iterative}, that do not need QPE. Of particular importance to this study is the MLQAE method by Suzuki et al.~\cite{suzuki2020amplitude}, that has many desirable properties. First and foremost, since MLQAE does not use QPE, it avoids controlled operations in QPE and instead uses maximum likelihood estimation as a post-processing method. Another advantage is that it does not need ancilla qubits, which are necessary to achieve the desired accuracy in canonical QAE (CQAE). Finally, it is parallelizable because the queries can be executed simultaneously before they are post-processed using maximum likelihood estimation.  

In this paper, we study the numerical implementation of quantum Monte Carlo integration (see Sec \ref{sec:MCI}), whose building blocks are QAE and Grover’s search algorithm, \cite{grover1996fast} and its generalization, quantum amplitude amplification (QAA)~\cite{brassard2002quantum}. The focus of the study is the investigation of methods for QAE that are appropriate for NISQ devices. To this end, we compare and contrast the performance of MLQAE with CQAE from a practical standpoint. We implement these quantum algorithms in Qiskit \cite{qiskit}, an open source software suite for near-term quantum computing, and apply the transpiler in Qiskit to optimize the implementation on each target quantum device. This is followed by a comparative analysis of execution of these algorithms on IBMQ devices.



\subsection{Monte Carlo Method for Integration}
\label{sec:MCI}
Monte Carlo (MC) methods are a wide class of computational techniques that make use of random number sampling.
In this section, we review hit-or-miss Monte Carlo Integration (MCI) to estimate a definite integral numerically.

Given a domain $\mathscr{D} \subset \mathbb{R}^n$ with two real valued functions, $f,g : \mathscr{D} \subset \mathbb{R}^n \rightarrow \mathbb{R}$ (typically $g \equiv 0$) and $f > g$, consider the definite integral of $f - g$.  Hit-or-miss MCI starts by generating samples uniformly from $\mathscr{D} \times [a_{n+1}, b_{n+1}]$ where $a_{n+1} \leq \textrm{min}(f-g)$ and $b_{n+1} \geq \textrm{max}(f-g)$.  A sample $X = (x_1, x_2, \cdots, x_n, x_{n+1}) \in \mathscr{D} \times [a_{n+1}, b_{n+1}]$ is considered a hit if it satisfies $x_{n+1} < f(x_1, x_2, \cdots, x_n)$ and $x_{n+1} > g(x_1, x_2, \cdots, x_n)$.
This method can be formulated as follows:

\begin{equation*}
\label{eq:MCI}
\frac{\idotsint_{\mathscr{D}} (f-g) d \textbf{x}^n} {\textrm{Volume} (\mathscr{D} \times [a_{n+1}, b_{n+1}]))} \approx \frac{\# \textrm{of hits}}{\# \textrm{of samples}}.
\end{equation*}

\begin{figure}[h]
\centering
  \subfloat{%
    \includegraphics[width=.40\textwidth]{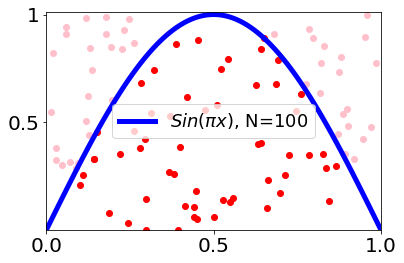}}
  \subfloat{%
    \includegraphics[width=.40\textwidth]{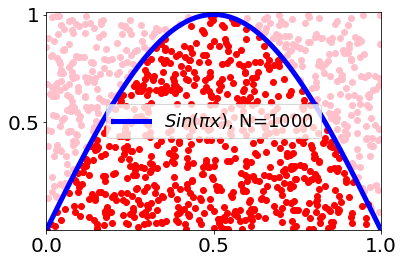}}\\
  \caption{A pictorial representation of Monte Carlo integration for $\sin( \pi x )$ on the interval $[0,1]$ using $100$ ($57$ sample points fall inside the region bounded by the curves, i.e, $57$ hits) and $1000$ ($640$ hits) uniformly sampled points, on the left and right hand side plots.}\label{fig:mc_example}
\end{figure}

The main advantage of MCI is that the error is solely a function of the number of the samples taken and so the basic process does not depend on the dimension of the function and thus can be scaled up easily, both in terms of effort and complexity, to higher dimensions. Deterministic numerical integration methods, such as the trapezoidal rule and Simpson's rule, have error bounds that increase exponentially as the dimension of the integral increases. All these factors make MCI an appealing method in higher dimension ($\gg 3$).
Figure \ref{fig:mc_example} shows a pictorial representation of MCI of $\int_{0}^{1} \sin( \pi x ) \, dx = \frac{2}{\pi} \approx 0.63662$. 
As shown in Fig. \ref{fig:mc_example}, we have $0.57$ from $100$ samples (left figure) and $0.64$ from $1000$ samples.

\subsection{Quantum Monte Carlo Integration}

Equation~(\ref{eq:MCI}) shows that MCI is evaluated by counting the number of hits when the volume of $\mathscr{D'} = \mathscr{D} \times [a_{n+1}, b_{n+1}]$, is easy to compute. If $\mathscr{D'}$ is complicated, then it can be divided into finite sub-domains that are simpler and the same procedure can be followed on each of the subdomains. MCI basically counts the number of good states ($\{ \textbf{x} ~|~ \chi(\textbf{x}) = 1  \}$) of the boolean function, $\chi : \mathscr{D'} \subset \mathbb{R}^{n+1} \rightarrow \{0, 1\}$, which depends on $f, g$ and domain $\mathscr{D'}$ in Eq.~(\ref{eq:MCI}).

One way of implementing MCI on a quantum computer is by counting the solutions of Grover's search algorithm. In fact, CQAE implements MCI by using the generalized Grover's algorithm (Quantum Amplitude Amplification) and QPE \cite{brassard2002quantum}.


\subsection{Notations}

We follow Dirac's Bra-Ket notation for qubit representation and arithmetic, such as the inner product and the tensor product.
For a multiple qubit system, we use the consecutive binary number strings with the most significant qubit located on the left and the least significant qubit on the right.
For example, we have $\ket{0} \otimes \ket{0} \otimes \ket{0} = \ket{0} \ket{0} \ket{0} = \ket{000}$ in the binary representation and $\ket{011} = \ket{3}$ and $\ket{110} = \ket{6}$ in the decimal representation of the computational basis.
In the decimal representation, the number of qubits $n$ is denoted by a subscript as in $\ket{0}_n$.
The dimension of a square matrix is also denoted by a subscript.
For example, $\mathbb{I}_{n}$ denotes the $n\times n$ identity matrix.
In quantum circuit diagrams, the top and the bottom qubits represent the least and the most significant qubits, respectively. The three Pauli matrices (Pauli gates when they are used in quantum circuits) $X$, $Y$ and $Z$ are as follows:

\begin{equation*}
\label{eq:pauli}
    X = \begin{pmatrix} 0 & 1\\ 1 & 0\end{pmatrix}, ~~
    Y = \begin{pmatrix} 0 & -i\\ i & 0\end{pmatrix}, ~~
    Z = \begin{pmatrix} 1 & 0\\ 0 & -1\end{pmatrix}.
\end{equation*}

\noindent The Hadamard matrix, H = $\frac{1}{\sqrt{2}} \begin{pmatrix} 1 & 1\\ 1 & -1\end{pmatrix}$, is a unitary matrix that is also Hermitian, so it is its own inverse.


\section{Quantum Amplitude Estimation Algorithms}


First, we briefly review quantum amplitude amplification and amplitude estimation algorithms \cite{brassard2002quantum}. The amplitude amplification is a generalization of Grover's search algorithm \cite{grover1996fast}, without the loss of the quadratic quantum speedup over its classical counterpart.

Suppose we have a boolean function $f$ and a unitary operator $\mathcal{A}$, where $f$ maps the good states to $1$ and the bad states to $0$ on domain $\mathscr{D}$ such that $| \mathscr{D} | = N = 2^n$, and $\mathcal{A}$ acts on $n+1$ qubits such that
\begin{equation}
\label{eq:qae_a}
    \ket{\Psi} = \mathcal{A} \ket{0}_n \ket{0} = \sqrt{1-a} \ket{\psi_0}_n \ket{0} + \sqrt{a} \ket{\psi_1}_n \ket{1},
\end{equation}
\noindent where the good state is $\ket{\psi_1}_n$ with $|\ket{\psi_1}_n| = k$, the bad state is $\ket{\psi_0}_n$, and $a = k / N \in [0,1]$ is unknown.
The job of the quantum amplitude estimation algorithm is to find $a$ approximately.
The algorithm complexity is measured by the number of quantum queries to the operator $\mathcal{A}$.

To achieve the quantum speedup, instead of measuring the last qubit of $\ket{\Psi} = \mathcal{A} \ket{0}_n \ket{0}$ directly, $\ket{\Psi}$ is first amplified by the following unitary operator $\textbf{Q}$:

\begin{equation}
\label{eq:qae_def_q}
    \textbf{Q} = \mathcal{A} \textbf{S}_0 \mathcal{A}^{-1} \textbf{S}_{\chi},
\end{equation}

\noindent where $\textbf{S}_0 = \mathbb{I}_{n+1} - 2 \ket{0}_{n+1} \bra{0}_{n+1}$ and
$\textbf{S}_{\chi} = ( \bigotimes\limits^{n} \mathbb{I}_2 ) \otimes Z$.
$\textbf{S}_{\chi}$ puts a negative sign to the good state $\ket{\psi_1}_n \ket{1}$ and does nothing to the bad state $\ket{\psi_0}_n \ket{0}$.
Let us define a parameter $\theta \in [0, \pi/2]$ so that $\sin^2 \theta = a$. With this, we can rewrite Eq.~(\ref{eq:qae_a}) as:
\begin{equation}
\label{eq:qae_theta}
    \ket{\Psi} = \mathcal{A} \ket{0}_n \ket{0} = \cos \theta \ket{\psi_0}_n \ket{0} + \sin \theta \ket{\psi_1}_n \ket{1}.
\end{equation}

\noindent By applying $\textbf{Q}$ (amplitude amplification operator) repeatedly $m$ times on $\ket{\Psi}$, we get 

\begin{equation}
\label{eq:qae_qm}
    \textbf{Q}^m \ket{\Psi} = \cos ((2m+1) \theta) \ket{\psi_0}_n \ket{0} + \sin ((2m+1) \theta) \ket{\psi_1}_n \ket{1}.
\end{equation}

\noindent From Eqs.~(\ref{eq:qae_a}) and (\ref{eq:qae_theta}), it can be observed that the measurement after applying $\textbf{Q}^m$ on $\mathcal{A} \ket{0}_n \ket{0}$ shows a quadratically larger probability of obtaining the good state (provided $\theta$ is sufficiently small so that $ (2m+1) \theta < \frac{\pi}{2}$) than measuring $\mathcal{A} \ket{0}_n \ket{0}$ directly \cite{brassard2002quantum}.


CQAE \cite{brassard2002quantum} estimates $\theta$ in Eq.~(\ref{eq:qae_theta}) by QPE which includes the inverse QFT. QPE is implemented by the controlled operation on the oracle queries. 
Hence, it needs a number of multi-controlled operations, which are further decomposed into many basis gates.
When this algorithm is implemented on NISQ devices, the accuracy is strongly limited by the connectivity between qubits because all the ancilla registers need connectivity with the target register and a sufficiently large number of ancillae are needed to ensure desired accuracy of the estimation.   
On the other hand, MLQAE \cite{suzuki2020amplitude} is implemented by post-processing the result of the quantum computation (and measuring the process without QPE) using maximum likelihood estimation.
Since different levels of amplification (depending on power of $Q$) can be executed independently, the algorithm is parallelizable. 
The implementation aspects of both CQAE and MLQAE will be discussed in the following sections.



\section{Implementation}
\label{sec:implementation}

For our study, a two-qubit domain is used to implement CQAE and MLQAE algorithms, and the solution is fixed to be $\ket{01}$ out of the four possible values, $\ket{00}$, $\ket{01}$, $\ket{10}$, and $\ket{11}$. The algorithms are implemented in Qiskit \cite{qiskit}, and optimized for IBM quantum devices using the transpiler in Qiskit.

\subsection{Implementation}

\begin{figure}[hbt]
\centering
  \subfloat[$\mathcal{A}$]{%
    \includegraphics[width=.25\textwidth]{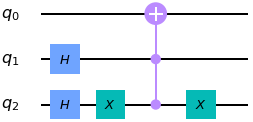}}
  \qquad\qquad\qquad\qquad\qquad
  \subfloat[$\mathcal{A}^{-1}$]{%
    \includegraphics[width=.25\textwidth]{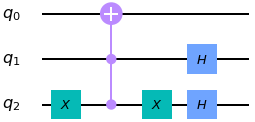}
  }\\
  \caption{Quantum circuit implementation of $\mathcal{A}$ from Eq.~(\ref{eq:qae_a}) in Qiskit.}\label{fig:A}
\end{figure}

\begin{figure} [ht]
\begin{center}
\includegraphics[width=0.99\linewidth]{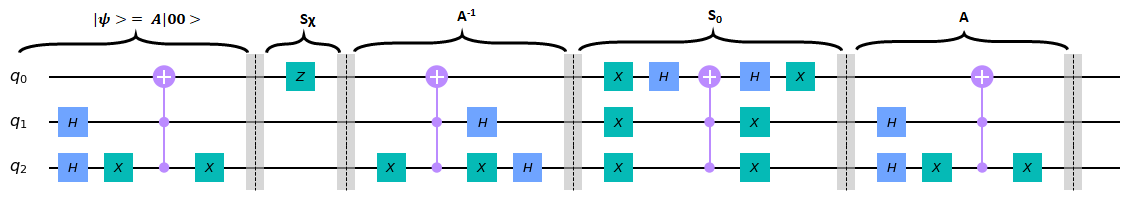}
\end{center}
\caption{Quantum circuit implementation of $\mathcal{A}$ and $\textbf{Q}$ from Eqs.~(\ref{eq:qae_a}) and~(\ref{eq:qae_def_q}), in Qiskit.}
\label{fig:Q} 
\end{figure}

The quantum circuit implementation of Eqs.~(\ref{eq:qae_a}) and~(\ref{eq:qae_def_q}) with a two-qubit domain in Qiskit are described in Figs.~\ref{fig:A} and~\ref{fig:Q}, respectively.
Figure~\ref{fig:A} also includes the quantum circuit of $\mathcal{A}^{-1}$ as it is used in Eq.~(\ref{eq:qae_def_q}).

\subsubsection{Canonical Quantum Amplitude Estimation}

\begin{figure} [htb]
\begin{center}
\includegraphics[width=0.99\linewidth]{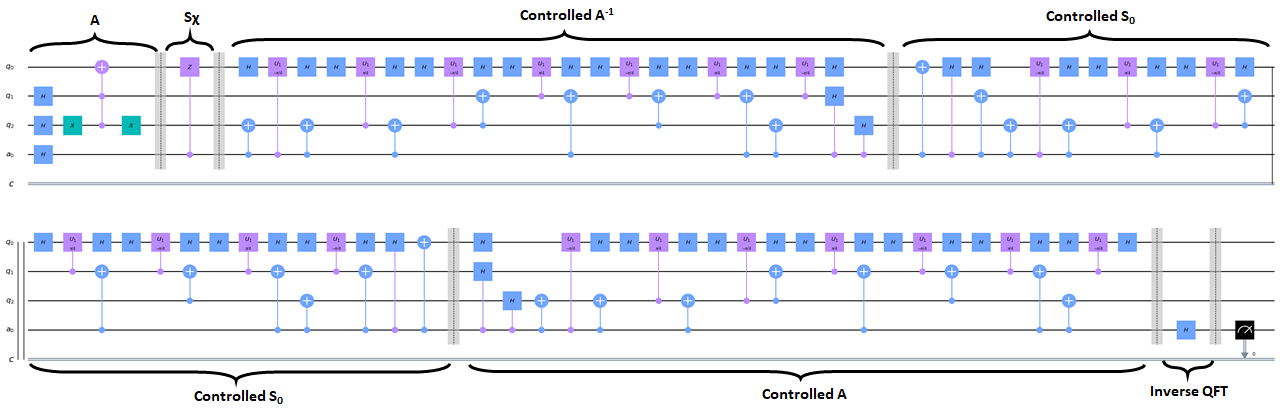}
\end{center}
\caption[example] 
{\label{fig:cae_2q_1m_sim}A CQAE quantum circuit implementation of the two-qubit domain with $m=1$. The circuit depth here is $78$.}
\end{figure} 

\begin{figure} [htb]
\begin{center}
\includegraphics[width=0.99\linewidth]{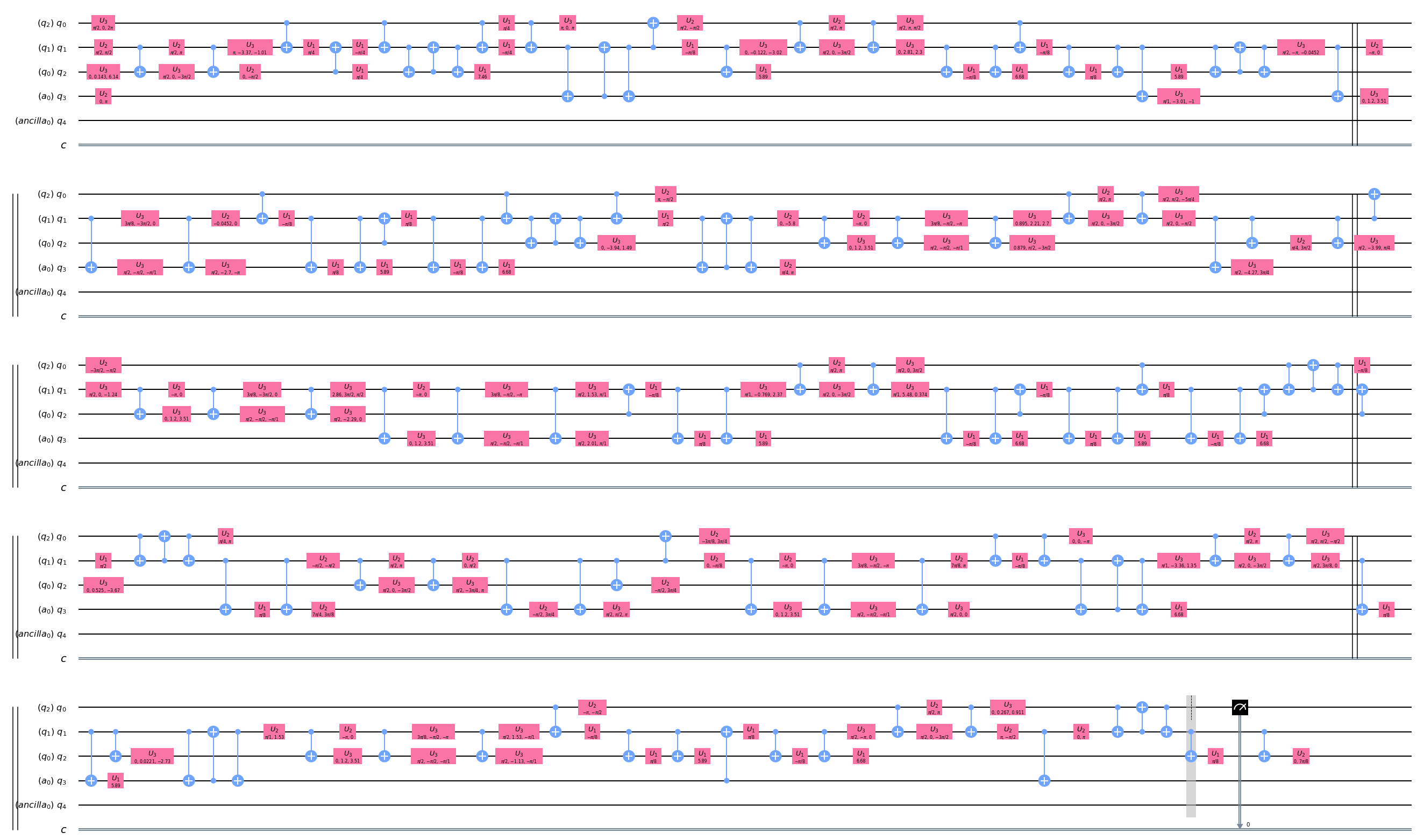}
\end{center}
\caption{The optimization of the two-qubit canonical QAE circuit using the transpiler in Qiskit on IBMQ VIGO. The circuit depth of this optimized circuit is $195$.}
\label{fig:cae_trans3_vigo} 
\end{figure} 

CQAE needs additional ancilla qubits to read out the amplitude.
The number of ancilla qubits is represented by $m$ in this paper.
Figures~\ref{fig:cae_2q_1m_sim} and~\ref{fig:cae_2q_3m_sim} show the quantum circuits with $m$ = $1$ and $3$, respectively.
Since the possible measured angle with $m$ ancillae are $2 \pi i / 2^m$ for $i = 0$ to $2^m - 1$, having more ancilla qubits increases the accuracy of the readout. For example, the possible measurements for $m=1$ are $0$ and $\pi$. The quantum circuit in Fig.~\ref{fig:cae_2q_1m_sim} is a variation of the circuit in Fig.~\ref{fig:Q}. The former adds control on the $\textbf{Q}$ operator followed by the inverse QFT.

To run a circuit on a quantum device, each quantum gate in the circuit should be decomposed into the basis gates supported by the device.
Figure~\ref{fig:cae_trans3_vigo} shows the decomposed and optimized circuit for the IBMQ VIGO device. For optimization, the transpiler in Qiskit is used with the optimization level set at $3$, the highest possible value allowed in the package.
In spite of applying the optimization, the circuit depth increases by more than double because of decomposition of the gates into the basis gates, and the addition of SWAP gates to the circuit to supplement disconnection between the qubits.

For $m=3$, the possible measurements of $\theta$ are $0, \pi/4, \pi/2, \cdots, 7 \pi / 4$.
The quantum circuit depth increases from $78$ (Fig.~\ref{fig:cae_2q_1m_sim}) to $514$ (Fig.~\ref{fig:cae_2q_3m_sim}) when $m$ is increased from $1$ to $3$ for more accuracy.
In Fig.~\ref{fig:cae_2q_3m_sim}, $a_0$, $a_1$ and $a_2$ represent the ancilla qubits and the circuit is composed of $\mathcal{A}$, followed by $a_0$ controlled $\textbf{Q}$, followed by $a_1$ controlled $\textbf{Q}^2$ (two consecutive $a_1$ controlled $\textbf{Q}$), followed by $a_2$ controlled $\textbf{Q}^4$ (four consecutive $a_2$ controlled $\textbf{Q}$), and the inverse QFT at the end.
The main reasons that lead to a sharp increase in the circuit depth are (a) exponential increase of controlled $\textbf{Q}$ operations, which increase to $2^m - 1$ if $m$ ancillae are used, and (b) controlled operations imposed on Toffoli operations in $\mathcal{A}$, $\textbf{Q}$, and $\mathcal{A}^{-1},$ as shown in Fig.~\ref{fig:Q}.
The controlled Toffoli (ccc-NOT) are composed of a number of basic quantum gates as shown in Figs.~\ref{fig:Q} and~\ref{fig:cae_2q_1m_sim}.
From Fig.~\ref{fig:cae_2q_1m_sim} to Fig.~\ref{fig:cae_trans3_vigo} ($m=1$ case), there is a twofold increases in circuit depth upon its decomposition into the basis gates for the quantum device.
For $m=3$, the circuit in Fig.~\ref{fig:cae_2q_3m_sim}, with a depth of $514$, should be decomposed into the basis quantum gates. As a consequence, its depth will become more than $1000$! And for that reason, we do not show the decomposed circuit in this paper. The implementation of such a circuit is not feasible on current NISQ devices because of decoherence and device error.


\subsubsection{Maximum Likelihood Quantum Amplitude Estimation}

The two main benefits of MLQAE over CQAE are (a) MLQAE does not need ancilla qubits to read out the amplitude $\sqrt{a}$ of Eq.~(\ref{eq:qae_a}), and (b) parallel execution.
The analysis of error bounds and the number of queries (application of $\mathcal{A}$ and $\mathcal{A}^{-1}$) are discussed by Suzuki et al.~\cite{suzuki2020amplitude} and Aaronson et al.~\cite{aaronson2020quantum}.
Both CQAE and MLQAE need $O \bigl( \frac{1}{\epsilon} \bigl( \sqrt{\frac{K}{N}} \bigr)   \bigr)$ oracle queries to estimate $K$ good states in $N$ samples (domain) with error $\epsilon$ \cite{aaronson2020quantum}.

\begin{figure}[ht]
\centering
  \subfloat[$\mathcal{A}$, circuit depth $4$]{%
    \includegraphics[scale=0.5]{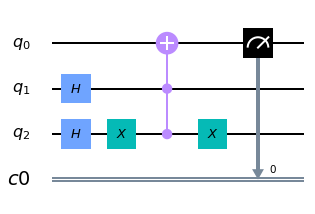}}
  \qquad\qquad    
  \subfloat[$\textbf{Q} \mathcal{A}$, circuit depth $15$]{%
    \includegraphics[scale=0.5]{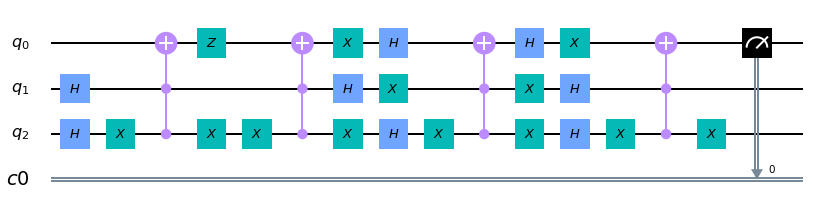}}\\
  \subfloat[$\textbf{Q}^2 \mathcal{A}$, circuit depth $26$]{%
    \includegraphics[scale=0.48]{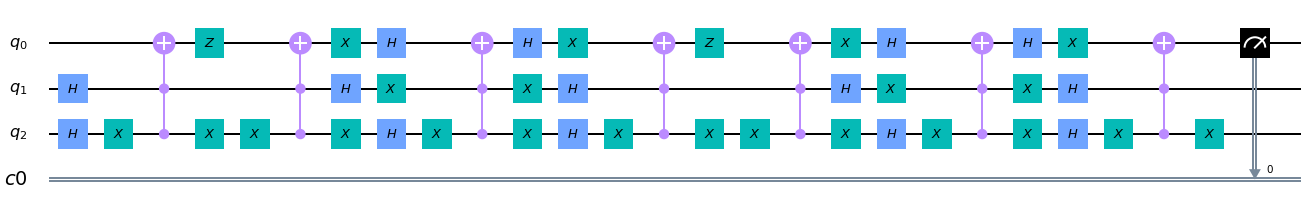}}\\
  \subfloat[$\textbf{Q}^4 \mathcal{A}$, circuit depth $48$]{%
    \includegraphics[scale=0.48]{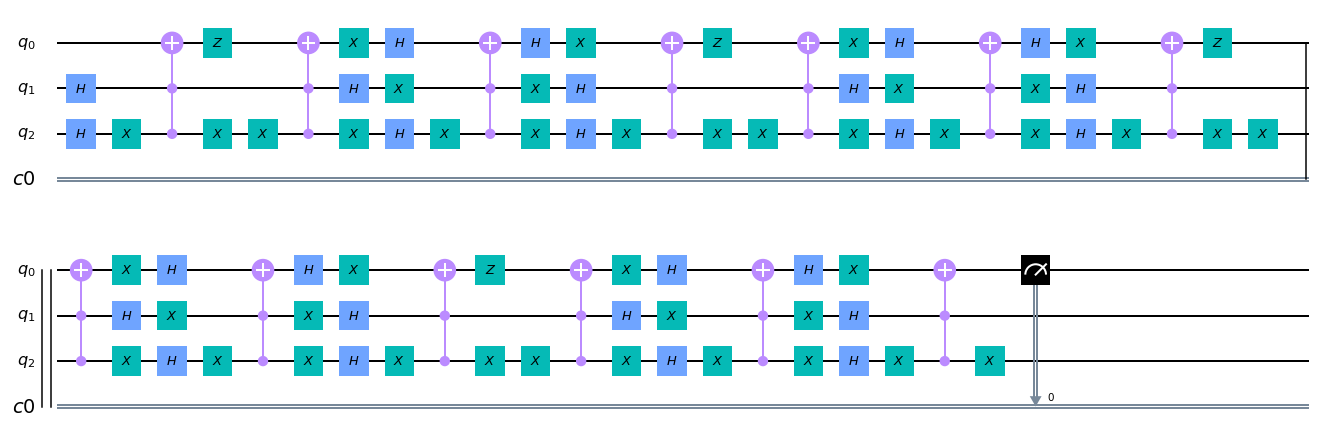}}\\
  \caption{An MLQAE quantum circuit implementation of two-qubit domain in Qiskit. We name circuits MLQAE$\left[ 0 \right]$, MLQAE$\left[ 1 \right]$, MLQAE$\left[ 2 \right]$, MLQAE$\left[ 3 \right]$ from (a) to (d), respectively.}
  \label{fig:mlae_2q_sim}
\end{figure}

Figure~\ref{fig:mlae_2q_sim} shows the quantum circuit implementation of MLQAE on a two-qubit domain.
When only one circuit (only $\mathcal{A}$) is used, then there is no quantum speed up~\cite{suzuki2020amplitude}. More circuits increase the accuracy of the estimation of the amplitude. Suzuki et al.~\cite{suzuki2020amplitude} discuss two options for circuit sequencing in MLQAE, linearly incremental sequence (LIS) and exponential incremental sequence (EIS), and suggest EIS as the asymptotically optimal choice.
Thus, we adapt the EIS which has exponential power of $\textbf{Q}$ from the second circuit. For example, the $n$-th circuit has $\textbf{Q}^{2^{n-2}}$ operator after $\mathcal{A}$ when we have $n > 1$.
Figure \ref{fig:mlae_2q_sim} shows four circuits of the MLQAE implementation and the last circuit (Fig.~\ref{fig:mlae_2q_sim} (d)) has $\textbf{Q}^4$ operator.

The number of $\textbf{Q}$ queries of MLQAE of $n$ circuits is equivalent to that of CQAE of $n-1$ ancilla qubits.
For example, when we have four circuits for MLQAE, the number of $\textbf{Q}$ queries is equivalent to that of CQAE having three ancilla qubits ($m=3$ case).
Since both CQAE and MLQAE have $O \bigl( \frac{1}{\epsilon} \bigl( \sqrt{\frac{K}{N}} \bigr)   \bigr)$ oracle queries, we can compare the two circuits of Figs.~\ref{fig:cae_2q_3m_sim} and~\ref{fig:mlae_2q_sim} with the same accuracy.
In this comparison, CQAE has a circuit depth of $514$ and MLQAE has a circuit depth of $93$ in total (sum of all circuit depths in Fig.~\ref{fig:mlae_2q_sim}).  The advantage offered by MLQAE with respect to decoherence due to its smaller circuit depth is even more pronounced because the circuits that make up the MLQAE run independently, and the measurements are post-processed.
For MLQAE, the noise is dominated by the last (longest) circuit due to decoherence and gate noise present in that circuit.
The circuits (a), (b), (c), and (d) in Fig.~\ref{fig:mlae_2q_sim} are named MLQAE$\left[ 0 \right]$, MLQAE$\left[ 1 \right]$, MLQAE$\left[ 2 \right]$, MLQAE$\left[ 3 \right]$, respectively.

\section{Results}
\label{sec:results}

In this section, we discuss the results obtained by running MLQAE on IBMQ simulator and IBMQ devices.
The simulator does not have quantum decoherence and gate errors.
Thus, we can regard the result from the simulator as perfect, meaning devoid of any quantum errors, although it does have errors and inaccuracies which arise from the algorithm, such as the discretization error.
Therefore, we can observe quantum noise or error from the implementation of the algorithm by comparing the results between the simulator and the devices.


\subsection{Maximum Likelihood Estimation}

The key idea of MLQAE is the post-processing of the measurements from each of the quantum circuits.
The likelihood functions and the resultant maximum likelihood function are defined as following in the domain $[0, \pi/2]$ for $\theta$:

\begin{equation}
\label{eq:lfunc}
    L_k (h_k ; \theta) = \lbrace \sin^2 ((2m_k+1) \theta) \rbrace^{h_k} \lbrace \cos^2 ((2m_k+1) \theta) \rbrace^{N - h_k},
\end{equation}

\begin{equation}
\label{eq:mlfunc}
    L (\vec{\textbf{h}} ; \theta) = \prod\limits_{k=0}^{m}  L_k (h_k ; \theta),
\end{equation}

\noindent where $m_k$, $h_k$, and $N$ are the power of $\textbf{Q}$, hit count of $1$, and number of shots for the $k$-th circuit, respectively, and $\vec{\textbf{h}} = (h_0, h_1, \cdots, h_m)$.
The maximum likelihood estimation estimates $\hat{\theta}$, which maximizes $L (\vec{\textbf{h}} ; \hat{\theta})$ in the domain.
But instead of $L (\vec{\textbf{h}} ; \theta)$, $\log L (\vec{\textbf{h}} ; \theta)$ is used to estimate $\hat{\theta}$ since the $\log$ function is monotonically increasing.
The Python code implementation is given in Appendix \ref{app:mle}.

\subsection{Simulator}

\begin{figure}[htb]
\centering
  \subfloat[$\mathcal{A}$, $248$ hits]{%
    \includegraphics[scale=0.65]{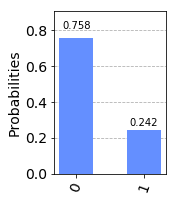}}
  \subfloat[$\textbf{Q} \mathcal{A}$, $1024$ hits]{%
    \includegraphics[scale=0.65]{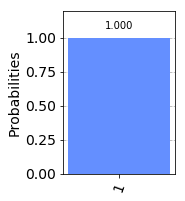}}
  \subfloat[$\textbf{Q}^2 \mathcal{A}$, $249$ hits]{%
    \includegraphics[scale=0.65]{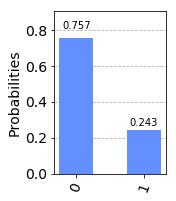}}
  \subfloat[$\textbf{Q}^4 \mathcal{A}$, $1024$ hits $\newline$]{%
    \includegraphics[scale=0.65]{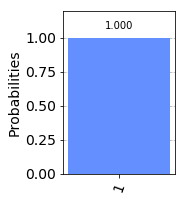}}
  \caption{The results of the MLQAE (Fig. \ref{fig:mlae_2q_sim}) on the IBMQ simulator with $1024$ shots.}\label{fig:mlae_2q_sim_result}
\end{figure}

The results from the IBMQ simulator are shown in Fig.~\ref{fig:mlae_2q_sim_result}.
Each histogram represents the result of its corresponding quantum circuit given in Fig.~\ref{fig:mlae_2q_sim}.
Since $a$ is $1/4$ and $\theta$ is $\pi/6$ in Eqs.~(\ref{eq:qae_a}) and~(\ref{eq:qae_theta}), in this setting, $3 \theta$, $5 \theta$, and $9 \theta$ are $\pi/2$, $5\pi/6$, and $3\pi/2$.
Therefore, circuits MLQAE$\left[ 0 \right]$, MLQAE$\left[ 1 \right]$, MLQAE$\left[ 2 \right]$, and MLQAE$\left[ 3 \right]$ will collapse to $\ket{1}$ with a probability of $0.25$, $1.00$, $0.25$, $1.00$, respectively, when they are measured in the states of Eq.~(\ref{eq:qae_qm}).
The results in Fig.~\ref{fig:mlae_2q_sim_result} are consistent with the analytical computation.

To estimate $\theta$ from the measured probability of each circuit, MLQAE uses the maximum likelihood estimation method~\cite{suzuki2020amplitude} (see Appendix~\ref{app:mle}). In this case, the estimated $\theta$ is $0.524$ (see Eqs.~(\ref{eq:qae_a}) and~(\ref{eq:qae_theta})).
Since the probability $a$, of measuring $\ket{1}$, is $\sin^2 (\theta)$, $a$ is $0.2504 \simeq 0.25$, as expected.

\subsection{IBM Quantum Devices}

\begin{figure} [ht]
\centering
  \subfloat[IBMQX2 $\newline$ U2 error (5.452e-4, 1.007e-3) $\newline$ CNOT error (1.434e-2, 2.689e-2)]{%
    \includegraphics[width=0.3\linewidth]{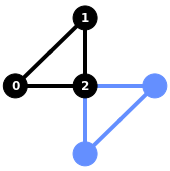}} \qquad\qquad\qquad\qquad
  \subfloat[IBMQ VIGO $\newline$ U2 error (3.632e-4, 6.552e-4) $\newline$ CNOT error (7.252e-3, 1.108e-2) $\newline$ ]{%
    \includegraphics[width=0.3\linewidth]{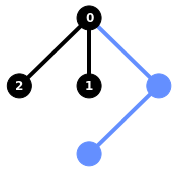}}
\caption{\label{fig:mlae_2q_ibmq_circuit_layout} The optimized layout of qubits for the circuits transpiled for IBMQX2 and IBMQ VIGO backends from Fig.~\ref{fig:mlae_2q_sim}(d), MLQAE$\left[ 3 \right]$. Errors shown are for a single qubit from the device calibration done on March 28, 2020. 
} 
\end{figure}

The optimizer (transpiler) maps logical qubits to device qubits optimally by considering qubit connectivity, and adds a SWAP gate if there is no physical connection between qubits.
Figure~\ref{fig:mlae_2q_ibmq_circuit_layout} shows the layout of how qubits, such as $q_0$, $q_1$, and $q_2$ in Fig.~\ref{fig:mlae_2q_sim}(d), are mapped to IBM quantum devices.
For example, on IBMQX2, $q_0$, $q_1$, and $q_2$ are mapped to $0$, $1$, and $2$ as can be seen in Fig.~\ref{fig:mlae_2q_ibmq_circuit_layout}. 

The optimized circuits of the MLQAE implementation (see Fig.~\ref{fig:mlae_2q_sim}) for IBMQX2 and IBMQ VIGO are shown in Figs.~\ref{fig:mlae_2q_ibmqx2_QA} and~\ref{fig:mlae_2q_vigo_QA}, respectively.
The main difference between the two devices is the qubit connectivity.
The three qubits on IBMQX2 have inter-connectivity with one another while the qubits $1$ and $2$ on IBMQ VIGO are not connected.
Thus, the optimized circuit for IBMQ VIGO, Fig~\ref{fig:mlae_2q_ibmqx2_QA} (b) in Appendix~\ref{app:circuits}, includes SWAP gates which are highlighted by red boxes.
Since SWAP gates are decomposed into three CNOT gates, optimized circuits on IBMQ VIGO have longer circuit depth than the circuits for IBMQX2, as shown in Figs.~\ref{fig:mlae_2q_ibmqx2_QA} and~\ref{fig:mlae_2q_vigo_QA} in Appendix~\ref{app:circuits}.

\begin{figure}[htb]
\centering
  \subfloat[$\mathcal{A}$, $468$ hits]{%
    \includegraphics[scale=0.65]{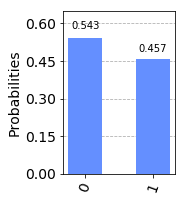}} 
  \subfloat[$\textbf{Q} \mathcal{A}$, $738$ hits]{%
    \includegraphics[scale=0.65]{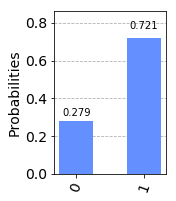}}
  \subfloat[$\textbf{Q}^2 \mathcal{A}$, $595$ hits]{%
    \includegraphics[scale=0.65]{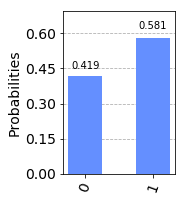}}
  \subfloat[$\textbf{Q}^4 \mathcal{A}$, $667$ hits $\newline$]{%
    \includegraphics[scale=0.65]{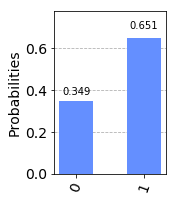}}\\
  \caption{The results of the MLQAE (Fig. \ref{fig:mlae_2q_ibmqx2_QA}) on IBMQX2 quantum device with $1024$ shots.}\label{fig:mlae_2q_ibmqx2_result}
\end{figure}

\begin{figure}[htb]
\centering
  \subfloat[$\mathcal{A}$, $274$ hits]{%
    \includegraphics[scale=0.65]{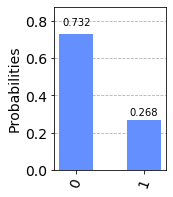}} 
  \subfloat[$\textbf{Q} \mathcal{A}$, $712$ hits]{%
    \includegraphics[scale=0.65]{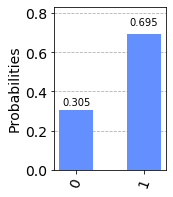}}
  \subfloat[$\textbf{Q}^2 \mathcal{A}$, $401$ hits]{%
    \includegraphics[scale=0.65]{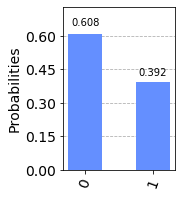}}
  \subfloat[$\textbf{Q}^4 \mathcal{A}$, $589$ hits $\newline$]{%
    \includegraphics[scale=0.65]{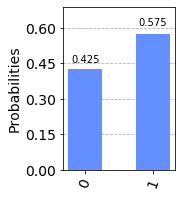}}\\
  \caption{The results of the MLQAE (Fig. \ref{fig:mlae_2q_vigo_QA}) on VIGO quantum device with $1024$ shots.}\label{fig:mlae_2q_vigo_result}
\end{figure}

Figures~\ref{fig:mlae_2q_ibmqx2_result} and~\ref{fig:mlae_2q_vigo_result} are histograms of measurements on quantum devices.
The results from IBMQ VIGO are more accurate than those from IBMQX2.
The result from the MLQAE[0] circuit on IBMQX2 (Fig.~\ref{fig:mlae_2q_ibmqx2_result} (a)) shows a relative error of $83.8\%$, whereas that value on IBMQ VIGO (Fig.~\ref{fig:mlae_2q_vigo_result} (a)) is only $7.2 \%$. This is because IBMQ VIGO has more accurate quantum device operations than IBMQX2, as depicted in Fig.~\ref{fig:mlae_2q_ibmq_circuit_layout}.
For the $\theta$-estimation problem (see Appendix \ref{app:mle}), IBMQX2 and IBMQ VIGO predict $\theta=0.795$ ($a=0.509$) and $\theta=0.780$ ($a=0.494$), respectively.
The former has $51.8 \%$ and $103.6 \%$ as the relative errors of $\theta$ and $a$ and those values for the latter are $49.0 \%$ and $97.6 \%$.
In contrast with the result from the MLQAE[0] circuit, there is not a significant difference between the overall accuracy of the two devices.
We hypothesize that IBMQ VIGO loses its accuracy because of the long circuit depths in MLQAE[2] and MLQAE[3].

\section{Conclusion}
\label{sec:discussion}

In this paper, we have implemented and discussed two quantum amplitude estimation algorithms on IBM quantum devices, in the broader context of Monte Carlo integration.
Even though CQAE~\cite{brassard2002quantum} is a monumental algorithm and a brilliant extension of Grover's algorithm~\cite{grover1996fast}, it is extremely challenging and in some cases infeasible to implement on NISQ devices, because of the required number of controlled operations acting on $\textbf{Q}$ operators in QPE.
Recently developed quantum amplitude estimation algorithms do not use QPE.  We have implemented one such algorithm, MLQAE, along with CQAE on IBM quantum devices and discussed their applicability on NISQ devices.
As discussed in Sec.~\ref{sec:results}, MLQAE is practical even on NISQ devices and has several advantages, including smaller circuit depth and parallel execution.  The advancements in QAE for quantum Monte Carlo integration rely on quantum amplitude amplification (QAA). 

As shown in Figs.~\ref{fig:Q} and~\ref{fig:mlae_2q_sim}, one of the crucial components of QAA is the $\textbf{S}_0$ operator, and it needs a multi-controlled NOT operator.
In general, when we have an $n$-qubit problem domain ($2^n$), we need $n$ controlled NOT gates for each $\textbf{Q}$ operator.
For example, when we have a four-qubit domain ($16$), the $\textbf{S}_0$ gate needs the following quantum gate,

\begin{equation*}
\label{eq:ccccnot}
    \Qcircuit @C=1.0em @R=0.2em @!R {
        \lstick{ {q}_{0}} & \targ     & \qw & \qw\\
        \lstick{ {q}_{1}} & \ctrl{-1} & \qw & \qw\\
        \lstick{ {q}_{2}} & \ctrl{-1} & \qw & \qw\\
        \lstick{ {q}_{3}} & \ctrl{-1} & \qw & \qw\\
        \lstick{ {q}_{4}} & \ctrl{-1} & \qw & \qw\\
    }
\end{equation*}

The implementation of the `advanced' mode of the above circuit in Qiskit is shown in Fig.~\ref{fig:ccccnot_sim}.
Even though the `basic' mode has shorter circuit depth, it needs two additional ancilla qubits as shown in Fig.~\ref{fig:ccccnot_sim_anc}.
The circuit decomposition of the `advanced' mode for IBMQX2 and IBMQ VIGO are shown in Fig.~\ref{fig:ccccnot_ibmqx2} and~\ref{fig:ccccnot_vigo}, respectively.
Since IBMQ VIGO has less connectivity than IBMQX2 (see Fig.~\ref{fig:mlae_2q_ibmq_circuit_layout}), the circuit for IBMQ VIGO has much longer depth ($181$) than that of IBMQX2 ($142$).
This example shows that the multi-controlled NOT gates have mainly two difficulties on NISQ devices.
The first is that a multi-controlled NOT gate should be decomposed into many basis gates as shown in Figs.~\ref{fig:ccccnot_ibmqx2} and~\ref{fig:ccccnot_vigo}.
The second difficulty is that all qubits which are involved in the gate should be connected. 
As shown in Fig.~\ref{fig:ccccnot_vigo}, the lack of connectivity between qubits are complemented by SWAP gates (decomposed into three CNOT gates), increasing decoherence through a longer circuit and increasing gate error.
Therefore, it is necessary to develop more efficient circuits for $\textbf{S}_0$, or better algorithms, to avoid the problems $\textbf{S}_0$ has in order to run scalable quantum Monte Carlo integration on NISQ devices.

\bibliography{report} 
\bibliographystyle{spiebib} 

\newpage

\appendix

\section{Maximum Likelihood Estimation Implementation}
\label{app:mle} 

$~\newline~$

\begin{lstlisting}[language={Python}, caption={The Maximum Likelihood Estimator Implementation}, label={list:after}]

import numpy as np
from scipy.optimize import brute

def MaximumLikelihoodEstmator(circuit_length, ones, zeros):

    grid = 20000
    epsilon = 1/grid  
    domain = [0.0 + epsilon, np.pi/2 - epsilon] # to avoid zero

    def logL(theta):
        fval = 0
        for i in range(circuit_length):
            if i==0:
                fval += 2 * ones[i] * np.log( np.absolute(np.sin(theta)) ) 
                fval += 2 * zeros[i] * np.log( np.absolute(np.cos(theta)) )
            else:
                fval += 2*ones[i]*np.log(np.absolute(np.sin((2*(2**(i-1))+1)*theta)))
                fval += 2*zeros[i]*np.log(np.absolute(np.cos((2*(2**(i-1))+1)*theta)))
        return -fval # to compute maximum

    return brute(logL, [domain], Ns=grid)[0]



ones_sim = [248, 1024, 249, 1024]
zeros_sim = [776, 0, 775, 0]

ones_ibmqx2 = [468, 738, 595, 667]
zeros_ibmqx2= [556, 286, 429, 357]

ones_vigo = [274, 712, 401, 589]
zeros_vigo = [750, 312, 623, 435]

est_theta_sim = MaximumLikelihoodEstmator(4, ones_sim, zeros_sim)
est_prob_sim = np.sin(est_theta_sim)**2

print('Estimated theta Simulator : ', est_theta_sim)
print('Estimated probability Simulator : ', est_prob_sim)

est_theta_ibmqx2 = MaximumLikelihoodEstmator(4, ones_ibmqx2, zeros_ibmqx2)
est_prob_ibmqx2 = np.sin(est_theta_ibmqx2)**2

print('Estimated theta IBMQX2 : ', est_theta_ibmqx2)
print('Estimated probability IBMQX2 : ', est_prob_ibmqx2)

est_theta_vigo = MaximumLikelihoodEstmator(4, ones_vigo, zeros_vigo)
est_prob_vigo = np.sin(est_theta_vigo)**2

print('Estimated theta IBMQ VIGO : ', est_theta_vigo)
print('Estimated probability IBMQ VIGO : ', est_prob_vigo)



\end{lstlisting}

\newpage

\section{Quantum Circuit Diagrams}
\label{app:circuits}

\begin{figure} [ht]
\begin{center}
\includegraphics[width=0.75\linewidth]{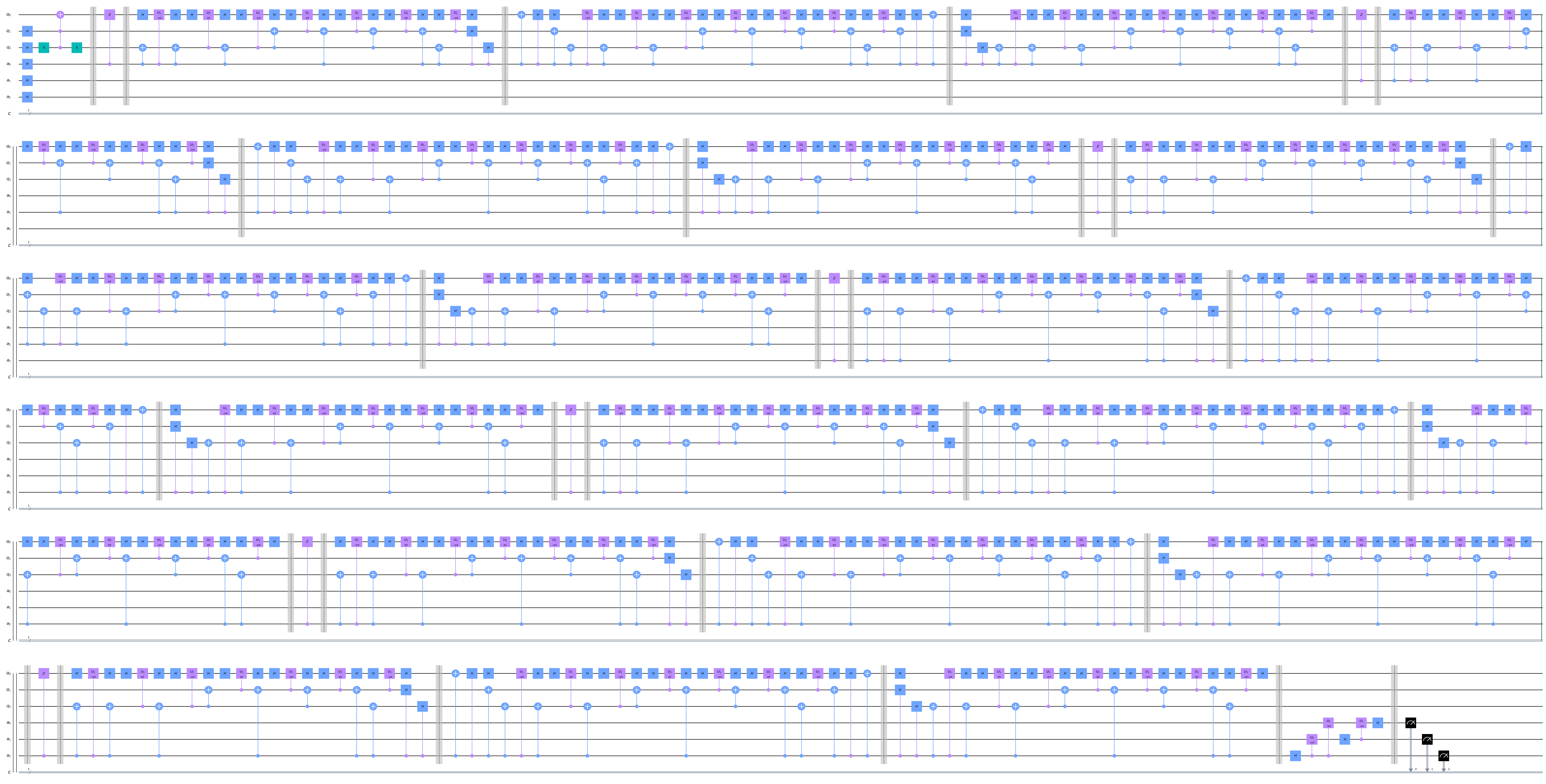}
\end{center}
\caption[example] 
{ \label{fig:cae_2q_3m_sim} 
A CQAE quantum circuit implementation of a two-qubit domain for $m=3$. The circuit depth is $514$.
}
\end{figure}

\begin{figure}[hbt]
\centering
  \subfloat[$\mathcal{A}$, circuit depth $10$ before the measurement]{%
    \includegraphics[scale=0.20]{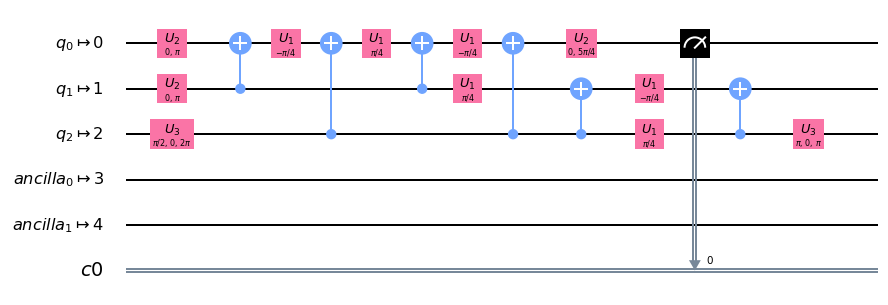}}
  \qquad\qquad    
  \subfloat[$\textbf{Q} \mathcal{A}$, circuit depth $42$ before the measurement]{%
    \includegraphics[scale=0.2]{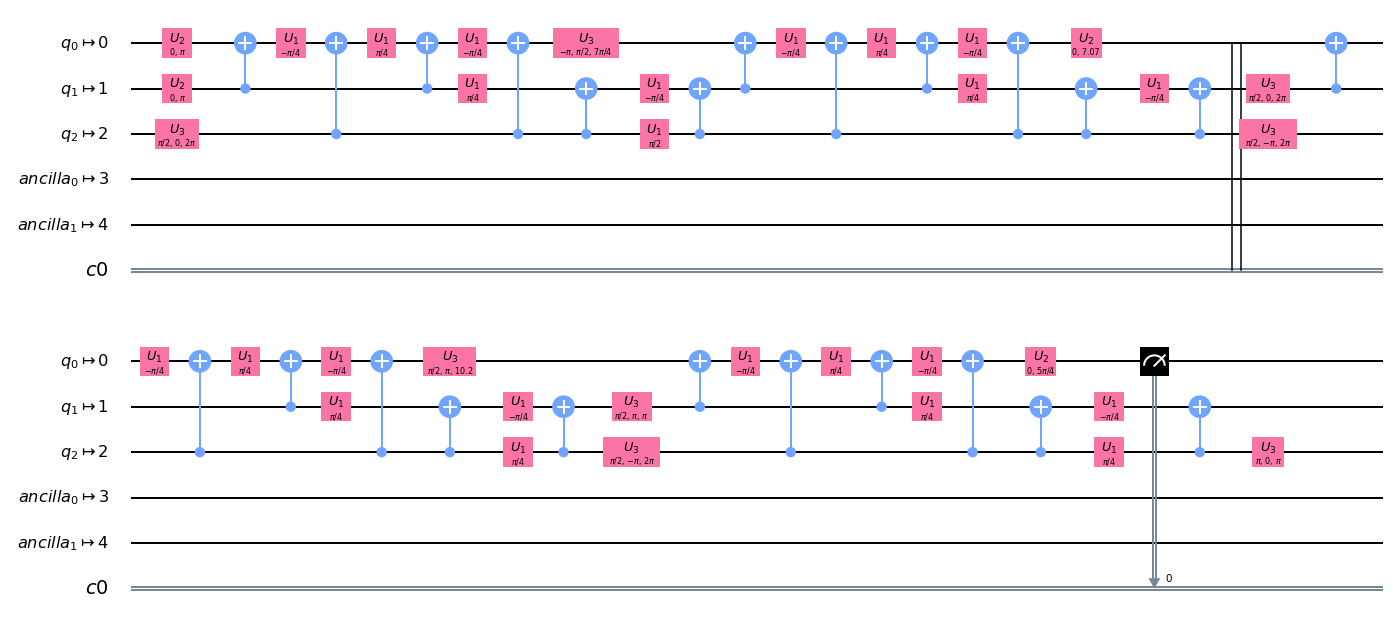}}\\
  \subfloat[$\textbf{Q}^2 \mathcal{A}$, circuit depth $74$ before the measurement$\newline$]{%
    \includegraphics[scale=0.25]{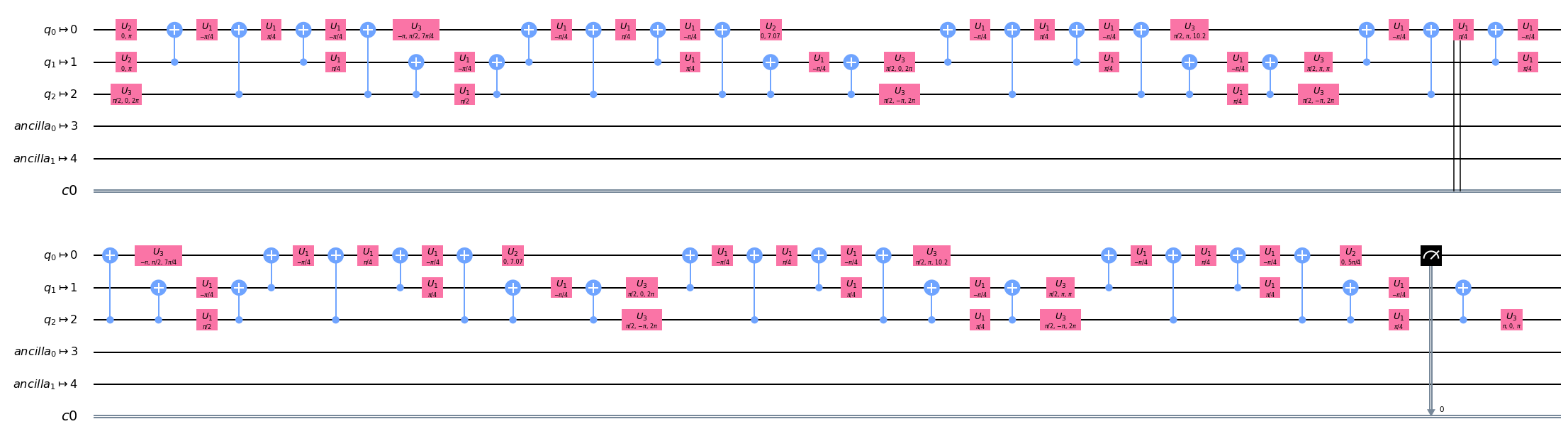}}\\
  \caption{The optimization of the two-qubit MLQAE circuit using the transpiler in Qiskit on IBMQX2 device. The optimization of Fig. \ref{fig:mlae_2q_sim} (d) is omitted because the circuit length is too long.}
  \label{fig:mlae_2q_ibmqx2_QA}
\end{figure}

\begin{figure}[ht]
\centering
  \subfloat[$\mathcal{A}$, circuit depth $13$ before the measurement]{%
    \includegraphics[scale=0.36]{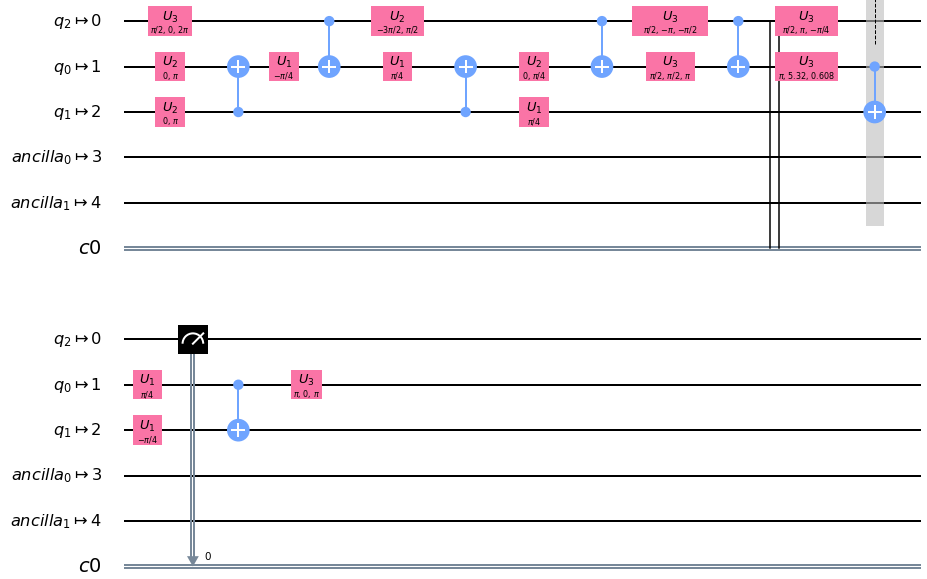}}
  \qquad\qquad    
  \subfloat[$\textbf{Q} \mathcal{A}$, circuit depth $72$ before the measurement]{%
    \includegraphics[scale=0.35]{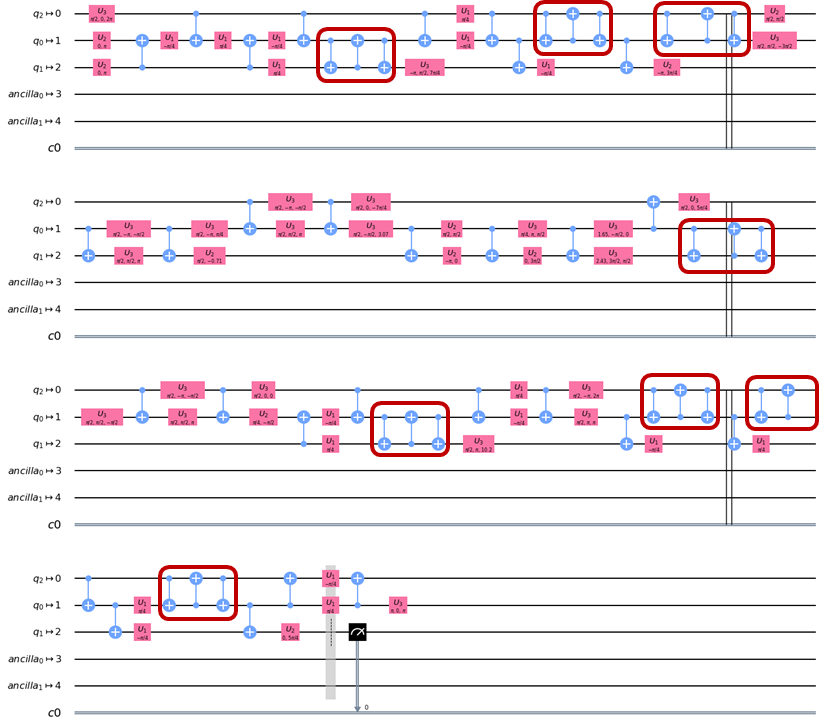}}\\
  \subfloat[$\textbf{Q}^2 \mathcal{A}$, circuit depth $121$ before the measurement$\newline$]{%
    \includegraphics[scale=0.422]{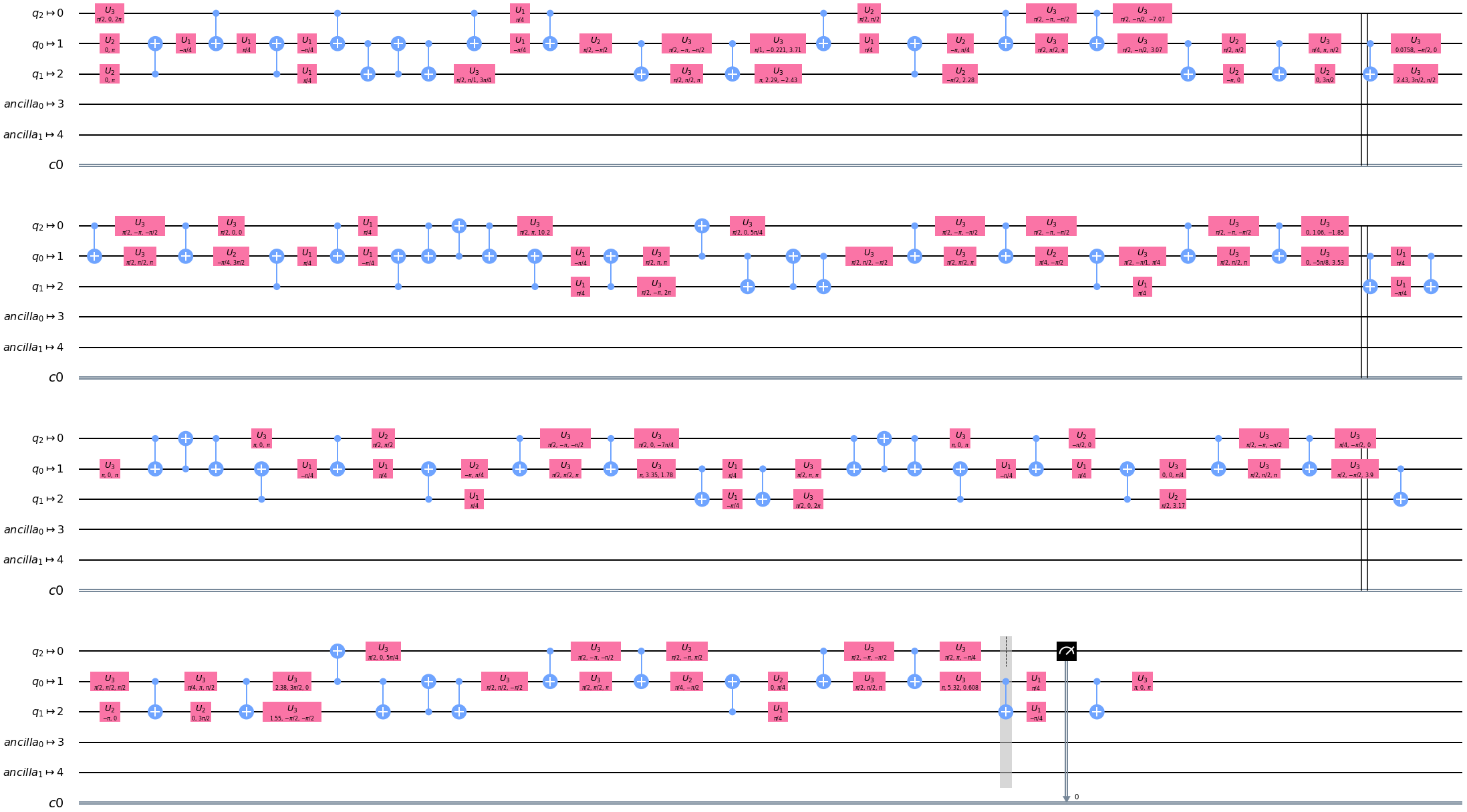}}\\
  \caption{The optimization of the two-qubit MLQAE circuit using the transpiler in Qiskit on IMBQ VIGO device. The optimization of Fig. \ref{fig:mlae_2q_sim} (d) is omitted because the circuit length is too long. In (b), SWAP gates are highlighted by red boxes.}
  \label{fig:mlae_2q_vigo_QA}
\end{figure}

\begin{figure} [ht]
\begin{center}
\includegraphics[width=0.99\linewidth]{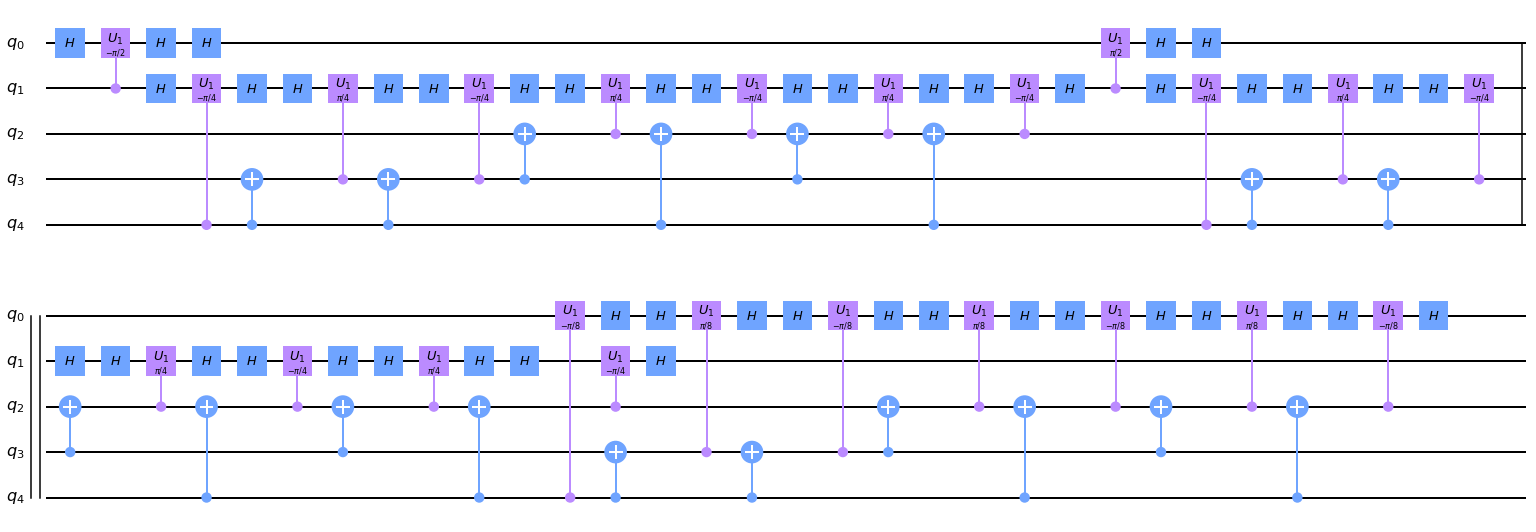}
\end{center}
\caption[example] 
{ \label{fig:ccccnot_sim} 
The cccc-NOT implementation in `advanced' mode in Qiskit. The circuit depth is $62$.
}
\end{figure}

\begin{figure} [ht]
\begin{center}
\includegraphics[width=0.99\linewidth]{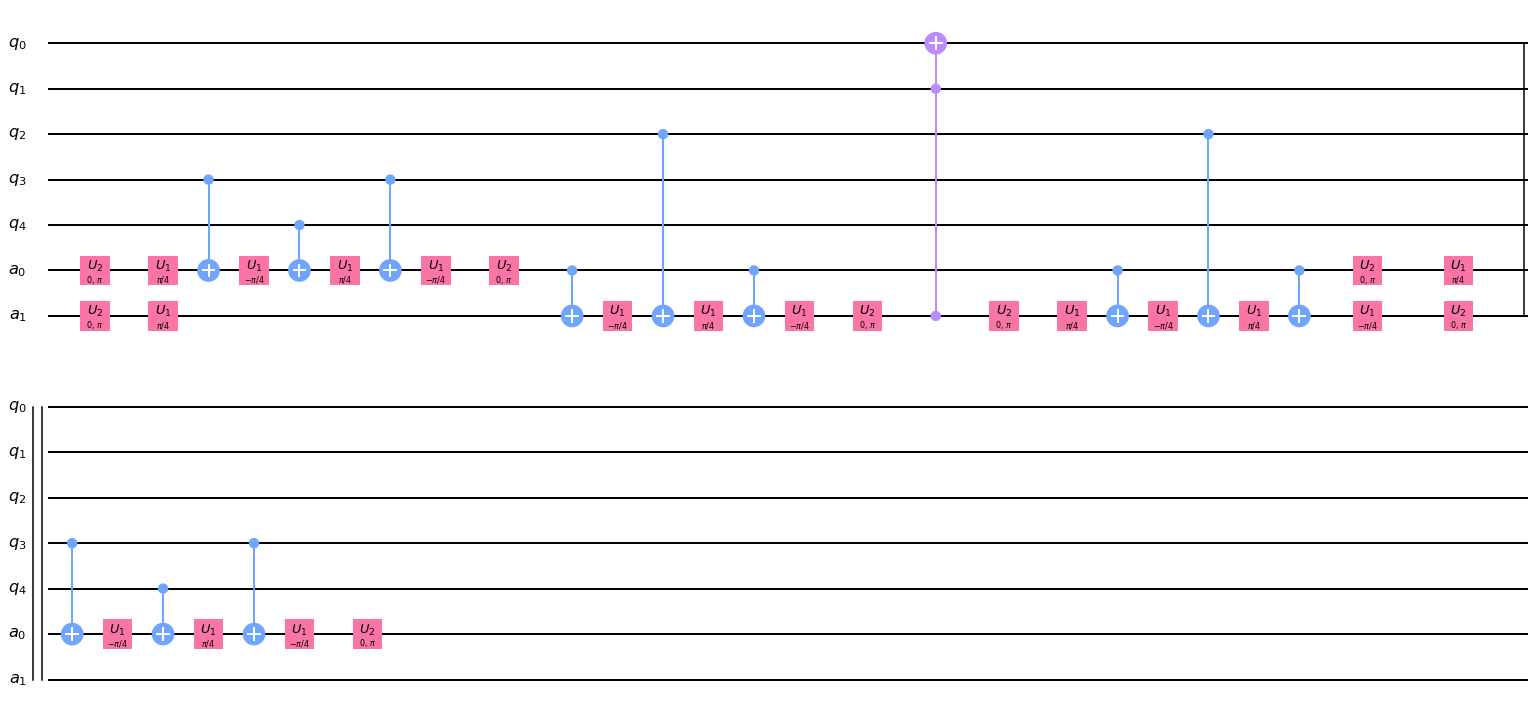}
\end{center}
\caption[example] 
{ \label{fig:ccccnot_sim_anc} 
The cccc-NOT implementation in `basic' mode in Qiskit. This mode needs two additional ancilla qubits. The circuit depth is $33$.
}
\end{figure}

\begin{figure} [ht]
\begin{center}
\includegraphics[width=0.99\linewidth]{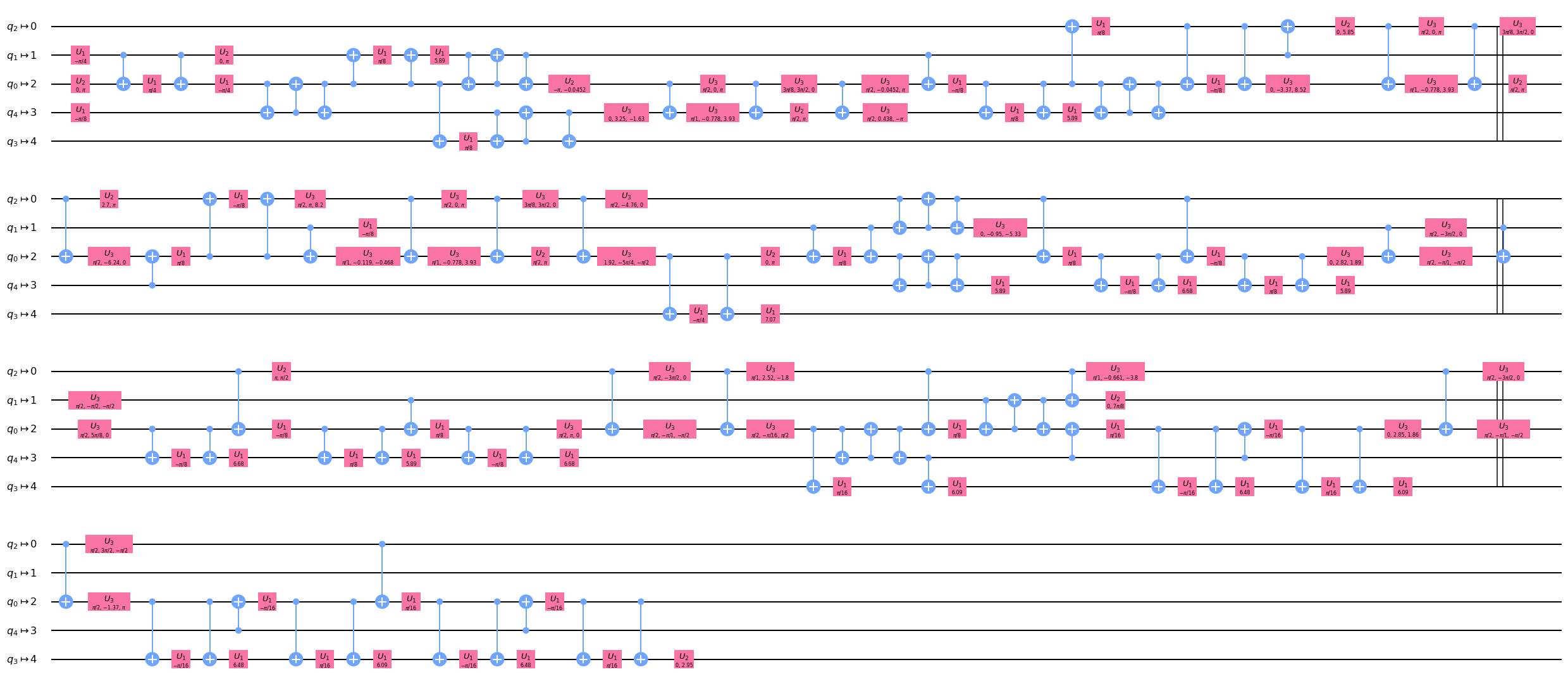}
\end{center}
\caption[example] 
{ \label{fig:ccccnot_ibmqx2} 
The cccc-NOT implementation on IBMQX2 in `advanced' mode. The circuit depth is $142$.
}
\end{figure} 

\begin{figure} [ht]
\begin{center}
\includegraphics[width=0.99\linewidth]{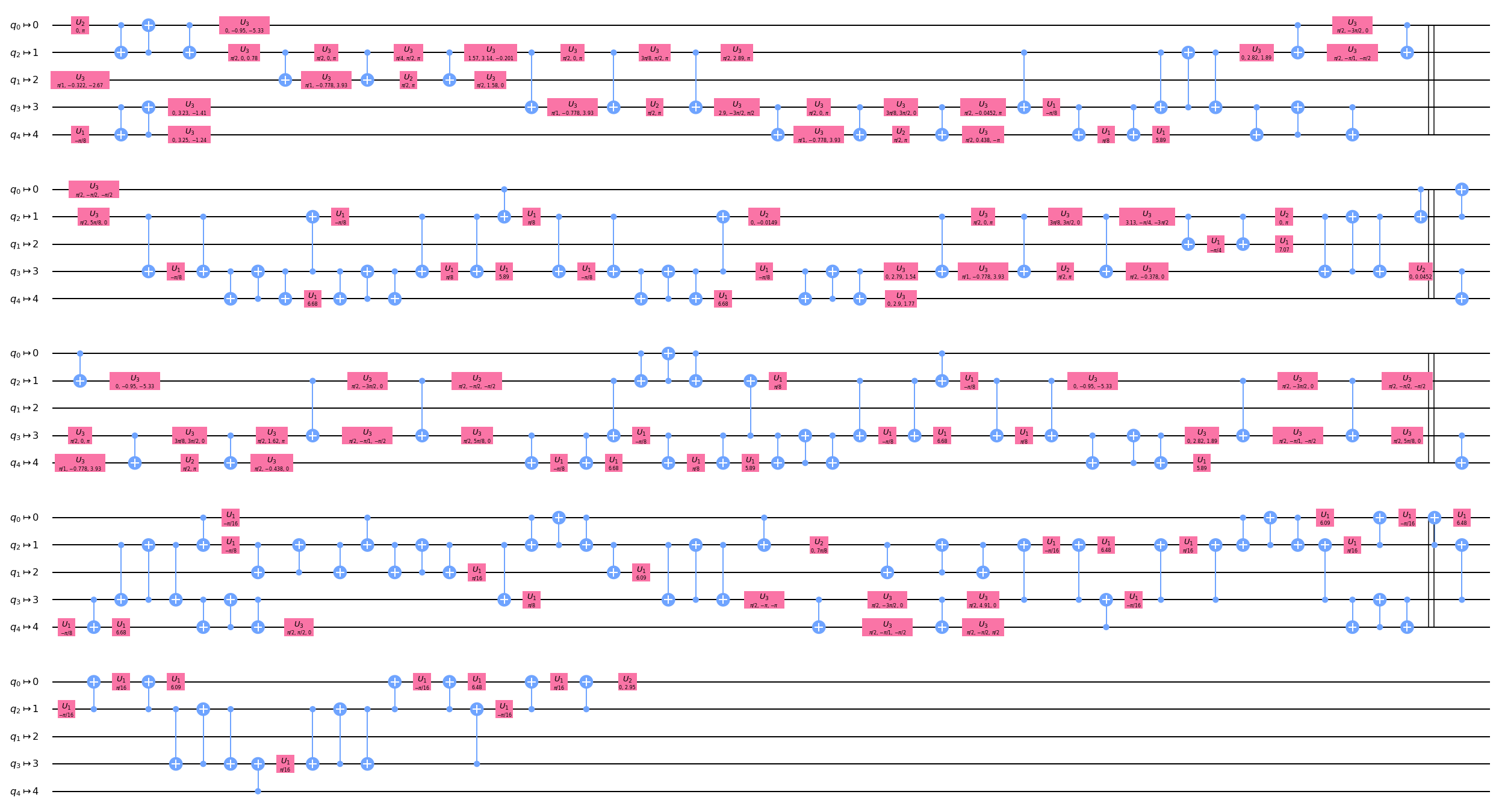}
\end{center}
\caption[example] 
{ \label{fig:ccccnot_vigo} 
The cccc-NOT implementation on IBMQ VIGO in `advanced' mode. The circuit depth is $181$.
}
\end{figure}

\end{document}